\documentclass[12pt]{article}

\usepackage[a4paper,left=30mm,right=20mm,top=20mm,bottom=20mm]{geometry}
\usepackage{graphics}
\usepackage{amsmath}
\usepackage{epsfig}
\usepackage{cite}
\usepackage{xcolor}
\usepackage[unicode]{hyperref}
\title{On the gluon shadowing effect in light and heavy nuclei at small $x$}

\author{A.V.~Lipatov$^{1}$, G.I.~Lykasov$^{2}$, M.A.~Malyshev$^{1,3}$}

\begin{document}

\maketitle

\begin{center}
{\it $^{1}$Skobeltsyn Institute of Nuclear Physics, Lomonosov Moscow State University, 119991 Moscow, Russia\\}
{\it $^{2}$Joint Institute for Nuclear Research, 141980 Dubna, Moscow region, Russia}\\
{\it $^{3}$Moscow Aviation Institute, 125993 Moscow, Russia}\\

\end{center}

\vspace{0.5cm}

\begin{center}

{\bf Abstract }

\end{center}

\indent

We propose new Transverse Momentum Dependent (TMD) 
gluon densities in nuclei.
Our method is based on a combination of the color dipole scattering formalism, 
within which we interpret the nuclear deep inelastic structure functions $F_2^{A}(x, Q^2)$ 
and well established property of geometrical scaling from nucleons to nuclei.
As an input, we employ the TMD gluon density in a proton
which provides a self-consistent simultaneous
description of the numerous HERA and LHC data
on $pp$, $ep$ and $\gamma p$ processes.
After fitting the relevant phenomenological parameters to the experimental data on the ratios
$F_2^{A}(x, Q^2)/F_2^{A^\prime}(x, Q^2)$
for several nuclear targets $A$ and $A^\prime$,
we derive the corresponding nuclear gluon distributions.
Then, we make predictions for gluon shadowing effects at small Bjorken $x$ in their
dependence of the mass number $A$.
A comparison with other approaches is given.

%We investigate the nuclear deep inelastic structure functions $F_2^{A}(x, Q^2)$
%within the color dipole formalism.
%Our analysis is based on the Transverse Momentum Dependent (TMD, or unitegrated)
%gluon density in a proton, which provides a self-consistent simultaneous
%description of the numerous HERA and LHC data, extended now to nuclei using the well established
%property of geometric scaling.
%By fitting some phenomenological parameters from the experimental
%data on ratios  $F_2^{A}(x, Q^2)/F_2^{A^\prime}(x, Q^2)$
%for several nuclear targets $A$ and $A^\prime$,
%we derive predictions for corresponding nuclear gluon distributions
%and, thus, for shadowing effects at small Bjorken variable $x$
%with respect to the mass number $A$.
%The comparison with other approaches is given.

\vspace{1.0cm}


%\newpage
\vspace{1.0cm}

\newpage

%\section{Introduction} \indent

It is well known that
a significant effect of nucleon interaction in the nucleus
appears in the deep inelastic scattering (DIS) of leptons on nuclei.
It demolishes the naive idea of the nucleus as a system of quasi-free nucleons\cite{EMCEffect, NuclReview-1, NuclReview-2,NuclReview-3}.
However, the standard concept of QCD factorization for $ep$ or $pp$ collisions,
where cross sections of arbitrary hard production processes are calculated
as a convolution of short-distance partonic cross sections and
parton (quark or gluon) distribution functions in a proton (PDFs),
is often extrapolated to $eA$ and $pA$ interactions.
In this case, proton PDFs are replaced by nuclear PDFs (nPDFs)
with keeping hard scattering cross sections the same\cite{EPPS21, nNNPDF3, nCTEQ}.
Therefore, detailed knowledge of nPDFs, in particular, the gluon distribution in nuclei,
is necessary for theoretical description of the
interaction dynamics in 
$eA$, $pA$ and $AA$ processes studied at
modern (LHC, RHIC) and future colliders (FCC-he, EiC, EicC, CEPC, NICA).

Influence of nuclear effects on PDFs meets a lot of interest from both
theoretical and experimental points of view.
Usually the nuclear modification factor, defined as a ratio of
per-nucleon structure functions in nuclei $A$ and deuteron\footnote{Or rather ratio of corresponding parton densities.}
$R = F_2^A(x,Q^2)/F_2^D(x,Q^2)$, is introduced
and its behaviour in the different $x$ ranges is
investigated.
So, there are shadowing, anti-shadowing, valence quarks and
Fermi motion dominance regions at $x \leq 0.1$, $0.1 \leq x \leq 0.3$, $0.3 \leq x \leq 0.7$ and $x \geq 0.7$, respectively.
The shadowing and anti-shadowing effects refer to $R < 1$ and $R > 1$ values,
whereas EMC effect and Fermi motion
refer to the slope of $R$ in the valence-dominant region and
rising of $R$ at larger $x$.
Different mechanisms are responsible for the different
nuclear effects. Unfortunately, up to now there is no commonly accepted framework to describe
the nuclear modification of PDFs in the
whole kinematical range.
Two main approaches are used by different groups.
In the first of them, nPDFs are extracted from a
global fit to nuclear data and their scale dependence is governed then
by standard Dokshitzer-Gribov-Lipatov-Altarelli-Parisi
(DGLAP) equations (see\cite{EPPS21, nNNPDF3, nCTEQ, nIMP}).
The second strategy is based on special nPDF models (see, for example,\cite{KulaginPetti-1, KulaginPetti-2} and references therein).
Nevertheless, quark and especially gluon distributions
in nuclei still have large uncertainties in the whole $x$
region due to shortage of experimental data and/or
limited kinematic coverage of the latter\cite{nPDFs-unc-1, nPDFs-unc-2, nPDFs-unc-3}.

Nowadays, approaches based on the
Transverse Momentum Dependent (TMD) parton
densities\footnote{The TMD parton densities are widely used in the Collins-Soper-Sterman approach\cite{CSS}
designed for semi-inclusive processes with a finite and non-zero ratio between the hard scale $\mu^2$ and
total center-of-mass energy $\sqrt{s}$. In the High Energy Factorization\cite{HighEnergyFactorization}, or $k_T$-factorization\cite{kt-factorization} approach,
developed for high energy hadronic collisions proceeding with a large momentum transfer and 
involving several hard scales, the TMD parton densities are often referred as unintegrated
parton distributions (uPDFs).}
have become quite popular in phenomenological analyses
of QCD processes (see, for example, reviews\cite{TMD-review, TMD-review-our} for more information).
In this Letter we investigate applicability of
a model\cite{LLM-2022, LLM-2024} 
%for TMD gluon density\cite{LLM-2022, LLM-2024} 
to the lepton-nucleus ($eA$) collisions.
Our study is inspired by the self-consistent simultaneous description
of numerous HERA and LHC data (in particular, data on the
proton structure function $F_2(x,Q^2)$, reduced cross sections $\sigma_r(x,Q^2)$ for the
electron-proton deep inelastic scattering at HERA and soft hadron production in $pp$ collisions at LHC)
achieved within this model.
Using the well established property of geometrical scaling\cite{GeometryScaling-1}
and an ansatz\cite{GeometryScaling-2},
we extend the proposed gluon density to nuclei
and study the structure function ratio
between various nuclei at low $x$
within the color dipole approach.
Such quantities are known to be a powerful tool 
to investigate the nuclear and nucleon structure. 
Performing a fit on available experimental data\cite{EMC-CCuSn,EMC-CCa1,EMC-CCa2,EMC-Cu,NMC-CCaovLi,NMC-HeCCaCCaovLi,NMC-LiC,NMC-BeAlCaFeSnPbovC,NMC-SnovC,SLAC-Fe,E665-Xe,E665-CCaPb},
we derive nuclear TMD gluon densities (nTMDs) at low $x$ ($x \leq 0.1$) for
several nuclei targets and then investigate corresponding
shadowing effects with respect to the mass number $A$.
The consideration
below continues the line of our studies\cite{LLM-2022, LLM-2024, LLM-FL, LLM-V+jet, LLM-InputUpdated}.
Similar calculations were done earlier\cite{GeometryScaling-2, nGBW}. However, 
these calculations based on the popular Golec-Biernat-W\"usthoff (GBW) model\cite{GBW-1, GBW-2},
which have some known difficulties in description of data (see, for example,\cite{GBW-problems1}). 
This is in a clear contrast with our input\cite{LLM-2022, LLM-2024}.

For the reader’s convenience, first we recall some important formulas.
So, in the color dipole formalism\cite{ColorDipoleModel-1,ColorDipoleModel-2,ColorDipoleModel-3, ColorDipoleModel-4,ColorDipoleModel-5}
the cross section of scattering between the transversely ($T$) and longitudinally ($L$) polarized
virtual photons $\gamma^*$ and proton or nucleus $A$ at small $x$ reads
\begin{gather}
  \sigma_{T, \, L}^{\gamma^* h}(x, Q^2) = \int d^2 {\mathbf r} \int\limits_0^1 dz |\Psi^{\gamma^*}_{T, \, L}(z,r,Q^2)|^2 \, \hat \sigma^h(x, r^2),
\label{eq-Sigma}
\end{gather}
\noindent
where the total cross section $\sigma^{\gamma^* h}(x, Q^2) = \sigma_{T}^{\gamma^* h}(x, Q^2) + \sigma_{L}^{\gamma^* h}(x, Q^2)$ with $h = p$ or $A$,
$\Psi^{\gamma^*}_{T, \, L}(z,r,Q^2)$ are the transverse and longitudinal spin-averaged
perturbatively calculated
wave functions\cite{ColorDipoleModel-3} for the
splitting of a
photon $\gamma^*$ having virtuality $Q^2$
into a %color
dipole $q\bar q$ with transverse size $r$,
$z$ is the quark longitudinal momentum fraction with respect
to the photon momentum $q$ (and $1 - z$ for antiquark), $x = Q^2/(W^2 + Q^2)$, $Q^2 = - q^2$, $W^2 = (p + q)^2$
with $p$ being the proton or nucleus momentum.
The squared photon wave functions read
\begin{gather}
  |\Psi^{\gamma^*}_{T}(z,r,Q^2)|^2 = {6 \alpha \over 4\pi^2} \sum_f e_f^2 \left\lbrace [z^2 + (1 - z)^2] \epsilon^2 K_1^2(\epsilon r) + m_f^2 K_0^2(\epsilon r) \right\rbrace, \nonumber \\
  |\Psi^{\gamma^*}_{L}(z,r,Q^2)|^2 = {6 \alpha \over 4\pi^2} \sum_f e_f^2 \left\lbrace 4 Q^2 z^2 (1 - z)^2 K_0^2(\epsilon r) \right\rbrace,
\label{eq-PhotonWaveFunctions}
\end{gather}
\noindent
where $\epsilon^2 = z(1 - z)Q^2 + m_f^2$, $K_0$ and $K_1$ are McDonald functions and
summation is performed over the quark flavors $f$.
The dipole cross section $\hat \sigma^h(x,r^2)$ contains all information about the
target $h$ and hard interaction. It is very difficult to calculate $\hat \sigma^h(x,r^2)$ {\it ab initio} and, therefore, usually it is 
modelled (see, for example,\cite{GBW-1, GBW-2, BGBK} and references therein).
In the case of proton, the dipole cross section $\hat \sigma^p(x,r^2)$
within the GBW approach
was proposed in the form\cite{GBW-1, GBW-2}
\begin{gather}
  \hat \sigma^p(x,r^2) = \sigma_0 \left\{ 1 - \exp\left[ - {r^2 \over R_0^2(x)} \right]\right\}, \quad R_0^2(x) = {1 \over Q_0^2} \left( x\over x_0 \right)^\lambda,
  \label{eq-DipoleCrossSectionGBW}	
\end{gather}
\noindent
where $\sigma_0$, $Q_0$, $x_0$ and $\lambda$ are free parameters and 
gluon saturation\cite{ColorDipoleModel-2, GBW-1, GBW-2} at small $x$ (see also\cite{Saturation}) is taken into account.
In the improved (BGK) model, 
$\hat \sigma^p(x,r^2)$ at large enough $\mu^2$ reads\cite{BGBK}
\begin{gather}
  \hat \sigma^p(x,r^2) = \sigma_0 \left\{ 1 - \exp\left[ - {\pi^2 r^2 \alpha_s(\mu_r^2) xg(x,\mu_r^2) \over 3 \sigma_0} \right]\right\},
  \label{eq-DipoleCrossSectionBGK}	
\end{gather}
\noindent
where $xg(x,\mu^2_r)$ is the gluon distribution function, 
$\mu_r^2 = C/r^2 + \mu_0^2$ with $C$ and $\mu_0^2$ 
being fitted to the experimental data.
In the leading $\ln 1/x$ limit,
which corresponds to the
approximation of
two gluon exchange
between the color dipole $q\bar q$ and proton debris,
there is a relation between the gluon
distribution function in a proton and dipole cross section:
\begin{gather}
  \hat \sigma^p(x,r^2) = {4 \pi^2 \alpha_s \over 3} \int {d{\mathbf k}_T^2 \over {\mathbf k}_T^2} \left[1 - J_0(|{\mathbf k}_T|r)\right] f_g^p(x, {\mathbf k}_T^2),
\label{eq-SigmaDipole}
\end{gather}
\noindent
where $J_0$ is the Bessel function of zeroth order and $\alpha_s$ is the fixed QCD coupling.
The gluon density $f_g^p(x, {\mathbf k}_T^2)$ depends on the transverse momentum ${\mathbf k}_T^2$
and is therefore called TMD (or unintegrated) one. So, from expressions~(\ref{eq-DipoleCrossSectionGBW}) and (\ref{eq-SigmaDipole})
one can immediately obtain
\begin{gather}
	f_g^p(x, {\mathbf k}_T^2) = {3 \sigma_0 \over 4 \pi^2} R_0^2(x) {\mathbf k}_T^2 \exp \left[ - R_0^2(x) {\mathbf k}_T^2 \right].
	\label{eq-GBWgluon}
\end{gather}	
\noindent
It is important that the so-called $x$-dependent saturation radius $R_0(x)$
involved into~(\ref{eq-DipoleCrossSectionGBW}) and (\ref{eq-GBWgluon}) determines the corresponding saturation scale $Q_s(x) = 1/R_0(x)$.
The latter regulates the partial saturation dynamics of gluon densities
at scales less than $Q_s(x)$: $f_g^p(x, {\mathbf k}_T^2) \simeq const \times \ln 1/x$\cite{ColorDipoleModel-2, GBW-1, GBW-2}.

In our previous paper\cite{LLM-2022} the TMD gluon density
in a proton at low scale $\mu^2 \simeq Q_0^2$ was proposed in another form (LLM):
\begin{gather}
  f_g^p(x, {\mathbf k}_T^2) = c_g (1 - x)^{b_g} \sum_{n = 1}^3  c_n \left(R_0(x) |{\mathbf k}_T| \right)^n \exp( - R_0(x)|{\mathbf k}_T|), \nonumber \\
  b_g = b_g(0) + {4 C_A \over \beta_0} \ln {\alpha_s(Q_0^2) \over \alpha_s({\mathbf k}_T^2)},
\label{eq-input}
\end{gather}
\noindent
where $C_A = N_c$, $\beta_0 = 11 - 2/3 N_f$ and $Q_0 = 2.2$~GeV.
All phenomenological parameters, namely, $c_g$, $x_0$, $\lambda$, $b_g(0)$ and $c_n$ were determined\cite{LLM-2022, LLM-InputUpdated, LLM-2024} from a fit to
numerous HERA and LHC data for processes extremely sensitive to the gluon content of a proton.
In particular, gluon density~(\ref{eq-input})
provides a self-consistent simultaneous
description of the HERA data on the proton structure function $F_2(x,Q^2)$, reduced cross section $\sigma_r(x,Q^2)$
for the electron-proton deep inelastic scattering at low $Q^2$ and soft hadron production in $pp$
collisions at the LHC
conditions\footnote{Being evolved according to the Catani-Ciafaloni-Fiorani-Marchesini equation\cite{CCFM}, the gluon density~(\ref{eq-input}) provides
good description of the LHC data on heavy flavor and Higgs production at different energies,
HERA data on the prompt photon production, proton structure functions $F_2^c(x,Q^2)$, $F_2^b(x,Q^2)$ and
longitudinal structure function $F_L(x, Q^2)$ in a wide range of $x$ and $Q^2$\cite{LLM-2022, LLM-2024,LLM-FL,LLM-V+jet, LLM-InputUpdated}.}.
Dynamics of the gluon saturation predicted by~(\ref{eq-input}) was specially
investigated\cite{LLM-2022, LLM-InputUpdated}. So, it was shown that expression~(\ref{eq-input}) leads to a smaller $Q_s$
than the GBW model at the same $x$.

\begin{table} %\scriptsize
	\label{table1}
	\begin{center}
		\begin{tabular}{c c c c}
			\hline
			\hline
			& & & \\
			Targets & Experiment & Reference & Number of data points \\
			& & & \\
			\hline
			& & &\\
			$^4$He$/$D & NMC & \cite{NMC-HeCCaCCaovLi} & $11$ \\
			$^6$Li$/$D & NMC & \cite{NMC-LiC} & $17$ \\
			$^{12}$C$/$D & NMC & \cite{NMC-HeCCaCCaovLi} & $11$ \\
			$^{12}$C$/$D & NMC & \cite{NMC-LiC} & $17$ \\
			$^{12}$C$/$D & EMC & \cite{EMC-CCuSn} & $3$ \\			
			$^{12}$C$/$D & EMC & \cite{EMC-CCa1} & $15$ \\			
			$^{12}$C$/$D & EMC & \cite{EMC-CCa2} & $9$ \\			
			$^{12}$C$/$D & E665 & \cite{E665-CCaPb} & $10$ \\			
			$^{40}$Ca$/$D & EMC & \cite{EMC-CCa1} & $15$ \\						
			$^{40}$Ca$/$D & EMC & \cite{EMC-CCa2} & $9$ \\						
			$^{40}$Ca$/$D & NMC & \cite{NMC-HeCCaCCaovLi} & $11$ \\						
			$^{40}$Ca$/$D & E665 & \cite{E665-CCaPb} & $10$ \\						
			$^{56}$Fe$/$D & SLAC & \cite{SLAC-Fe} & $1$ \\
			$^{64}$Cu$/$D & EMC & \cite{EMC-CCuSn} & $3$ \\								    
			$^{64}$Cu$/$D & EMC & \cite{EMC-Cu} & $4$ \\		    
			$^{119}$Sn$/$D & EMC & \cite{EMC-CCuSn} & $3$ \\
			$^{131}$Xe$/$D & E665 & \cite{E665-Xe} & $9$ \\
			$^{208}$Pb$/$D & EMC & \cite{E665-CCaPb} & $10$ \\	
			$^{12}$C$/$$^6$Li & NMC & \cite{NMC-CCaovLi} & $17$ \\		    		    		    		    		    	    		    		    		    
			$^{12}$C$/$$^6$Li & NMC & \cite{NMC-HeCCaCCaovLi} & $17$ \\		    		    		    		    		    	    		    		    		    
			$^{40}$Ca$/$$^6$Li & NMC & \cite{NMC-CCaovLi} & $8$ \\		    		    		    		    		    	    		    		    		    
			$^{40}$Ca$/$$^6$Li & NMC & \cite{NMC-HeCCaCCaovLi} & $8$ \\
			$^{9}$Be$/$$^{12}$C & NMC & \cite{NMC-BeAlCaFeSnPbovC} & $8$ \\
			$^{27}$Al$/$$^{12}$C & NMC & \cite{NMC-BeAlCaFeSnPbovC} & $8$ \\
			$^{40}$Ca$/$$^{12}$C & NMC & \cite{NMC-BeAlCaFeSnPbovC} & $8$ \\
			$^{56}$Fe$/$$^{12}$C & NMC & \cite{NMC-BeAlCaFeSnPbovC} & $8$ \\
			$^{208}$Pb$/$$^{12}$C & NMC & \cite{NMC-BeAlCaFeSnPbovC} & $8$ \\
			$^{119}$Sn$/$$^{12}$C & NMC & \cite{NMC-BeAlCaFeSnPbovC} & $8$ \\
			$^{119}$Sn$/$$^{12}$C & NMC & \cite{NMC-SnovC} & $83$ \\
			& & &\\
			\hline
			\hline
		\end{tabular}
	\end{center}
	\caption{List of nuclear data used in our fit. Number of data points at $x < 0.1$ are shown in last column.}
	\label{tbl:NuclearData}
\end{table}

An essential point of our derivation below is the so-called geometric scaling\cite{GeometryScaling-1},
which is a characteristic feature of low-$x$ data on the deep inelastic lepton-proton scattering at HERA.
In fact, the data on
$\sigma^{\gamma^* p}(x,Q^2)$ cross section over a wide range of $Q^2$
can be described\footnote{The geometric scaling naturally arises in the non-linear small-$x$ QCD evolution
equations. See, for example,\cite{BK, GeometryScalingFromNonLinearQCDEvolution-1, GeometryScalingFromNonLinearQCDEvolution-2, GeometryScalingFromNonLinearQCDEvolution-3}.}
by a single variable $\tau = Q^2/Q_s^2(x)$,
where all $x$ dependence is encoded in the saturation scale $Q_s^2(x) \sim x^{- \lambda}$
with $\lambda \sim 0.3$.
The same scaling effect was observed for nuclear cross sections $\sigma^{\gamma^* A}(x,Q^2)$\cite{GeometryScaling-2}.
It was shown that
\begin{gather}
  {\sigma^{\gamma^* A}(\tau) \over \pi R_A^2} = {\sigma^{\gamma^* p}(\tau) \over \pi R_p^2},
\label{eq-GeometryScalingA}
\end{gather}
\noindent
where $R_h$ is the radius of the hadronic target and $\tau = \tau_A \equiv Q^2/Q_{s,\,A}^2$ or $\tau = \tau_p \equiv Q^2/Q_{s}^2$, respectively.
So, the $A$-dependence of the ratio $\sigma^{\gamma^* h}(x, Q^2) / \pi R_h^2$
can be absorbed in the $A$-dependence of $Q_{s,\,A}(x)$.
For this dependence, an ansatz was proposed\cite{GeometryScaling-2}:
\begin{gather}
  Q^2_{s,\,A}(x) = Q^2_s(x) \left( {A \pi R_p^2 \over \pi R_A^2} \right)^{1\over \delta}
\label{eq-SaturationScaleA}
\end{gather}
\noindent
where $\delta$ and $\pi R_p^2$ are free parameters.
The nuclear gluon distribution could be obtained from the gluon
density in a proton after replacement $Q_s^2(x) \to Q_{s, \, A}^2(x)$\cite{GeometryScaling-2}
and applying the overall normalization factor $R_A^2/R_p^2$, which immediately comes from (\ref{eq-GeometryScalingA}).

Now %, using the ansatz~(\ref{eq-SaturationScaleA}), 
we can %easily 
extend the
gluon density~(\ref{eq-input}) to the nuclei and investigate structure function ratios between various nuclei and/or deutron.
As usual, these structure functions, $F_2^h(x, Q^2)$, are calculated as the
sum of the transverse and longitudinal ones:
\begin{gather}
 F_2^h(x, Q^2) = F_T^h(x, Q^2) + F_L^h(x, Q^2),
\label{eq-F2-1}
\end{gather}
\noindent
where $F_T^h(x, Q^2)$ and $F_L^h(x, Q^2)$
can be obtained within the color dipole approach at low $x$
using~(\ref{eq-Sigma}) --- (\ref{eq-SigmaDipole}) as:
\begin{gather}
  F_{T, \, L}^h(x, Q^2) = {Q^2 \over 4\pi^2 \alpha} \sigma_{T, \, L}^{\gamma^* h}(x, Q^2).
\end{gather}
\noindent
%{\it + some text here about fit performed by Maksim and other results}

The formulae listed above allow us to calculate the structure functions $F_2^A(x,Q^2)$ for various nuclei. 
As mentioned above, the scaling parameters $\pi R_p^2$ and $\delta$ are to be fitted from data. 
Since we are interested in the low $x$ region, where shadowing takes place and gluon distribution 
contributes dominantly, we have performed such a fit on structure function ratio data taken at $x<0.1$ by the EMC\cite{EMC-CCuSn, EMC-CCa1, EMC-CCa2, EMC-Cu}, 
NMC\cite{NMC-HeCCaCCaovLi, NMC-LiC, NMC-CCaovLi, NMC-BeAlCaFeSnPbovC, NMC-SnovC}, 
SLAC\cite{SLAC-Fe} and E665\cite{E665-CCaPb, E665-Xe} Collaborations. 
The measured nuclear targets involved into
our analysis are listed in Table~\ref{tbl:NuclearData}.
These data make the total of $333$ points.
To estimate the nucleus radius % in (\ref{eq-SaturationScaleA})
we use well-known parametrization applied earlier\cite{GeometryScaling-2} and another one\cite{RA-Textbook}, namely
\begin{equation}
	R_A = \left(1.12 A^{1/3} - 0.86 A^{-1/3}\right)\text{\,fm},
	\label{RA}
\end{equation}
\noindent 
and 
\begin{equation}
	R_A = \left( 1.12 A^{1/3} - 0.5 \right)\text{\,fm}.
	\label{RA-Kapitonov}
\end{equation}
The latter formula may have an advantage of a better agreement with the proton radius $R_p\sim 0.7$~fm obtained from 
fits based on the scaling approach~\cite{GeometryScaling-2}. Below we denote the fit obtained with (\ref{RA}) as 'Fit I' 
and with (\ref{RA-Kapitonov}) as 'Fit II'.
Note that we take the deutron to be identical to the proton, so that $R_D=R_p$ and $A_D=1$.

\begin{figure}
\begin{center}
\includegraphics[width=3.9cm]{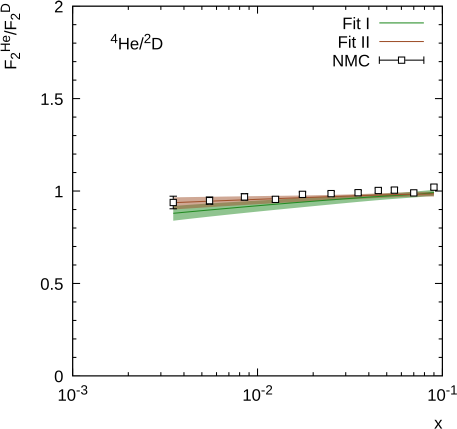}
\includegraphics[width=3.9cm]{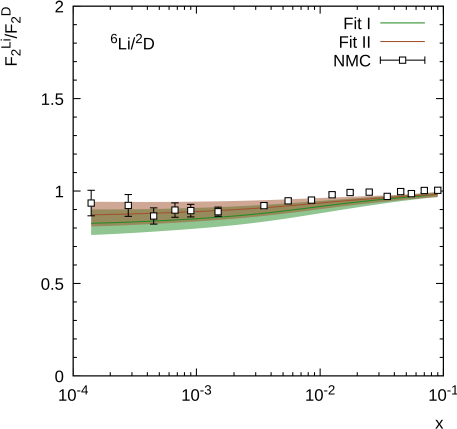}
\includegraphics[width=3.9cm]{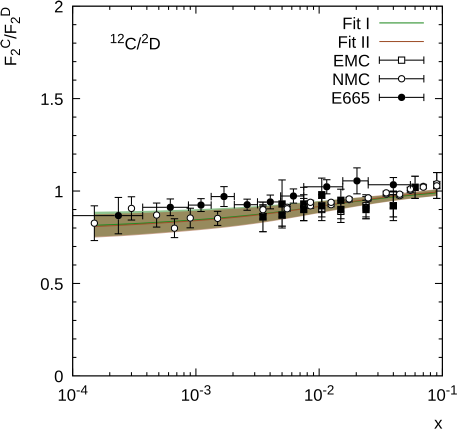} 
\includegraphics[width=3.9cm]{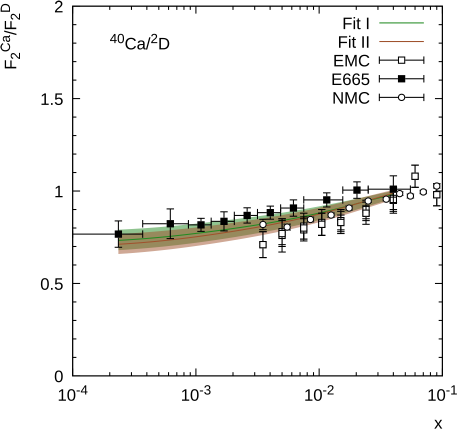}
\includegraphics[width=3.9cm]{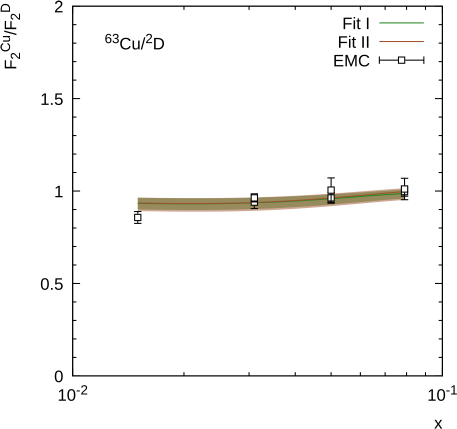}
\includegraphics[width=3.9cm]{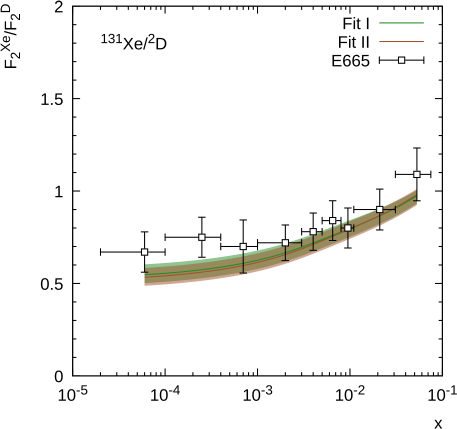}
\includegraphics[width=3.9cm]{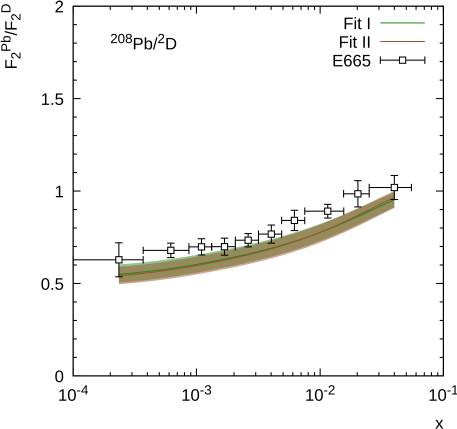}
\includegraphics[width=3.9cm]{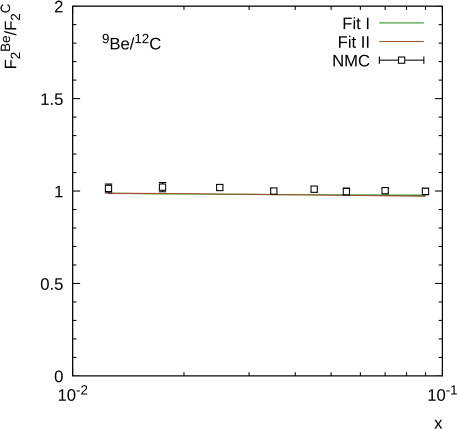}
\includegraphics[width=3.9cm]{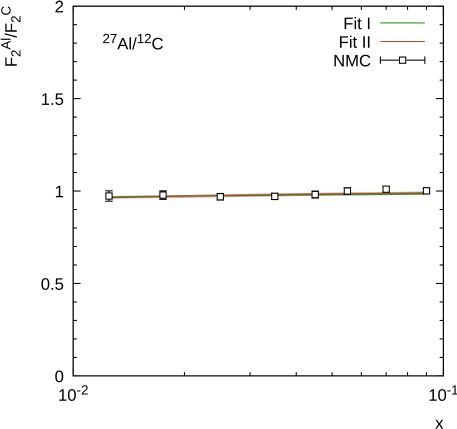}
\includegraphics[width=3.9cm]{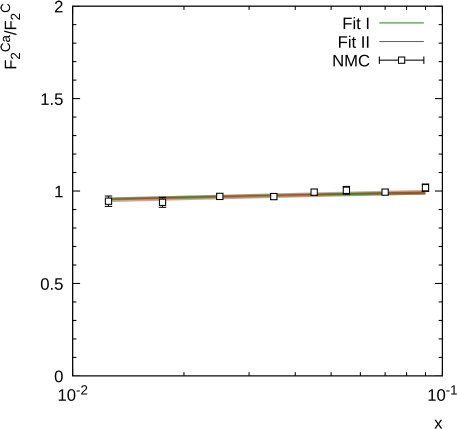}
\includegraphics[width=3.9cm]{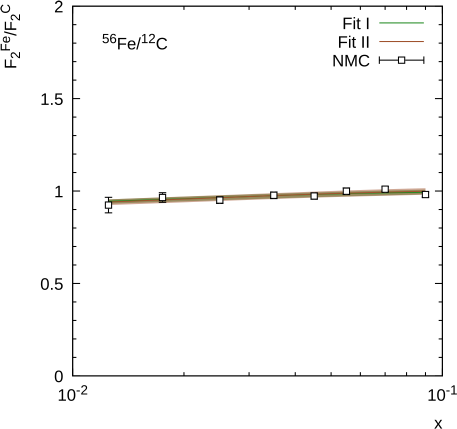}
\includegraphics[width=3.9cm]{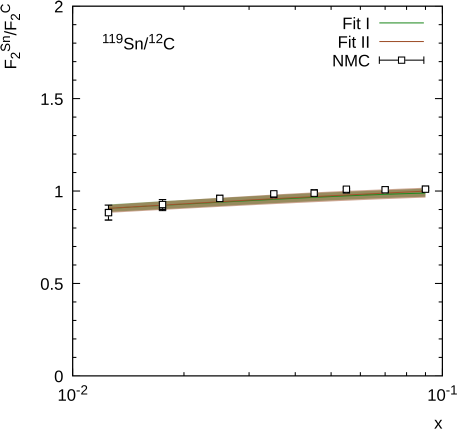}
\includegraphics[width=3.9cm]{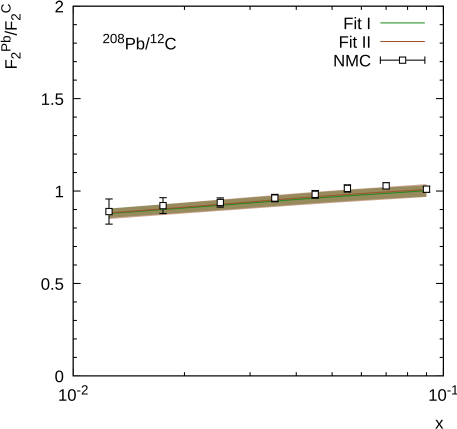}
\includegraphics[width=3.9cm]{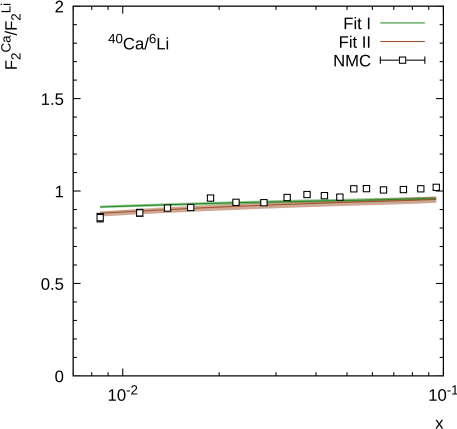}
\includegraphics[width=3.9cm]{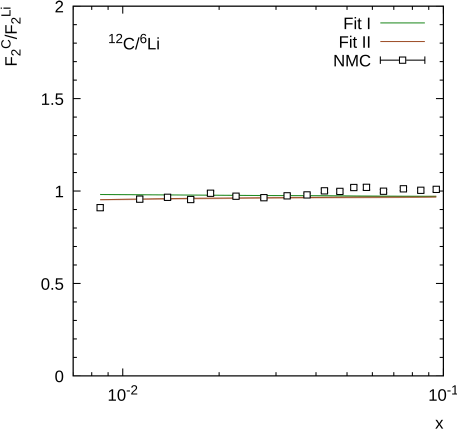}
\caption{The global fit results of structure function ratios
$F_2^{A}(x, Q^2)/F_2^{A^\prime}(x, Q^2)$ between different nuclei targets $A$ and $A^\prime$. 
Shaded bands represent the estimated uncertainties of our fit procedure.
Experimental data are from
NMC\cite{NMC-HeCCaCCaovLi, NMC-LiC, NMC-CCaovLi, NMC-BeAlCaFeSnPbovC, NMC-SnovC}, EMC\cite{EMC-CCuSn, EMC-CCa1, EMC-CCa2, EMC-Cu}, SLAC\cite{SLAC-Fe} and E665\cite{E665-CCaPb, E665-Xe}.}
\label{F2AF2A} 
\end{center}
\end{figure}

Our fits for $\pi R_p^2$ and $\delta$ parameters lead to 
$\pi R_p^2 = 1.74 \pm 0.20 $~fm$^2$, $\delta = 0.751 \pm 0.026$ with $\chi^2/n.d.f.= 2.28$ for Fit I 
and $\pi R_p^2 = 1.86 \pm 0.20$~fm$^2$, $\delta = 0.740^{+0.041}_{-0.021}$ with $\chi^2/n.d.f. = 2.19$ for Fit II.
%These values agree reasonably with the results\cite{GeometryScaling-2}, where the values
%$\pi R_p^2 = 1.55 \pm 0.02$~fm$^2$ and $\delta = 0.79 \pm 0.02$ have been obtained.
The experimental data involved into the fit procedures are compared with our predictions in Fig.~\ref{F2AF2A}.
It can be seen that the both fits allow to describe the various data quite well.
Some systematic underestimation can be seen for E665 data on $F_2^\text{Pb}/F_2^\text{D}$, $F_2^\text{Ca}/F_2^\text{D}$ and $F_2^\text{C}/F_2^\text{D}$, which, however, does not exceed two standard deviations at any point. Another source of discrepancy comes from NMC data on $F_2^\text{Sn}/F_2^\text{C}$. % at intermediate $Q^2$.
We find that Fit II leads to somewhat better agreement with the data,
especially the data on light nuclei ($^4$He, $^6$Li, $^{12}$C) at low $x$,
that is an immediate consequence of closer limit $R_A\to R_p\sim0.7$~fm for small $A$
achieved with~(\ref{RA-Kapitonov}).
Note that for the most of the used data the points are taken for different $Q^2$, 
so one should consider the lines depicted in Fig.~\ref{F2AF2A} as interpolations in both $x$ and $Q^2$.

\begin{figure}
	\begin{center}
		\includegraphics[width=3.9cm]{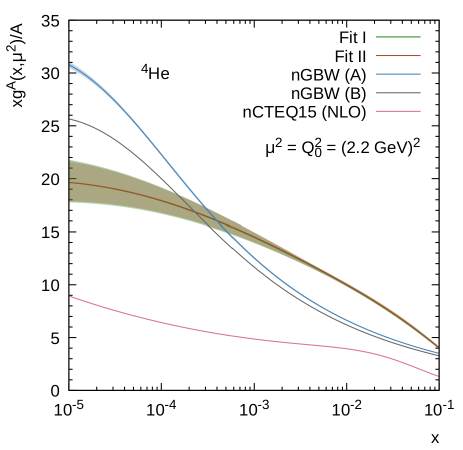}
		\includegraphics[width=3.9cm]{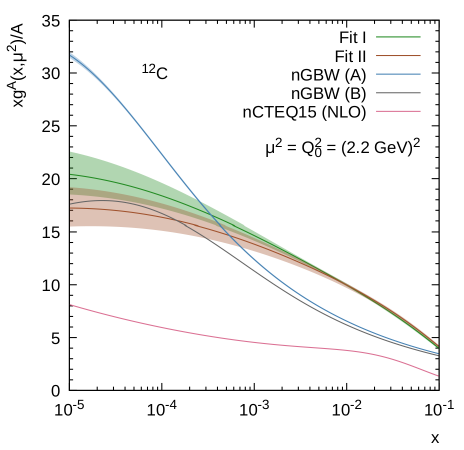}
		\includegraphics[width=3.9cm]{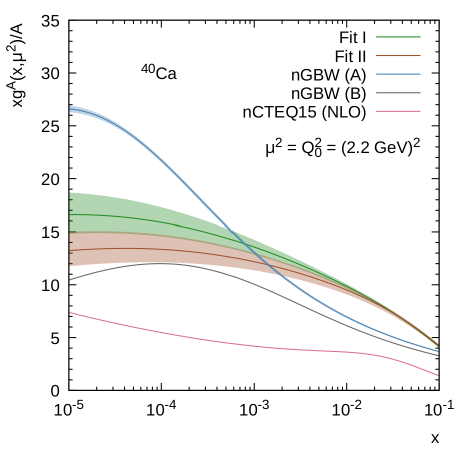}
		\includegraphics[width=3.9cm]{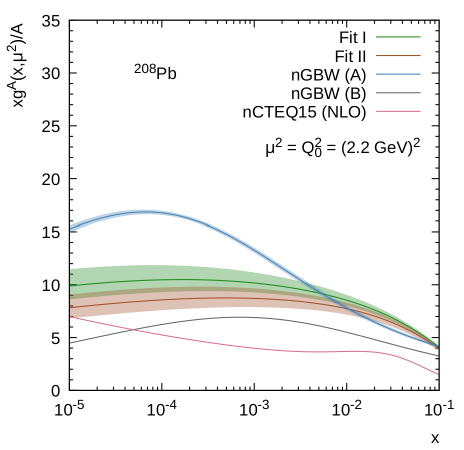} 
	\caption{Effective (integrated over ${\mathbf k}_T^2$) gluon densities in some light and heavy nuclei (namely, $^4{\rm He}$, $^{12}{\rm C}$, $^{40}{\rm Ca}$ and $^{208}{\rm Pb}$)
			calculated as functions of $x$ at $\mu^2 = Q_0^2$.
			For comparison we show here corresponding nPDFs obtained by the nCTEQ Collaboration\cite{nCTEQ}.}
		\label{fig3}
	\end{center}
\end{figure}

In Fig.~\ref{fig3} we present results for effective (integrated over ${\mathbf k}_T^2$ up to scale $\mu^2$) nuclear gluon distribution 
functions $xg^A(x, \mu^2)$ divided by $A$ for several nuclei as functions of the momentum fraction $x$
%taken at $\mu^2 = Q_0^2$ 
in the region of the fit, $x \leq 0.1$.
Such quantities, in principle, could be compared with ordinary nPDFs (see discussion below).
The upper limit of the integration is chosen to be equal to $\mu^2 = Q_0^2$,
where $Q_0$ is the starting scale of subsequent QCD evolution
applied to the LLM gluon density~(\ref{eq-input}), see\cite{LLM-2024}.
Our choice is motivated by the fact that all phenomenological parameters involved in~(\ref{eq-input}) were determined up to this scale.
Shaded bands represent the estimated uncertainties of our fit procedure.
As one can see, the difference between the Fit I and Fit II 
appears mainly at low $x$ and 
is more pronounced for heavier nuclei.
The obtained results can be compared with predictions\cite{GeometryScaling-2,nGBW}, %predictions
which are based on the GBW form of the gluon density~(\ref{eq-GBWgluon}).
So, the anzats~(\ref{eq-SaturationScaleA}) was applied
in the calculations\cite{GeometryScaling-2}
and another %often used 
condition, namely, $Q_{s,\,A}^2(x) = A^{1/3} Q_s^2(x)$,
was also tested\cite{nGBW}.
We label these predictions as nGBW (A) and nGBW (B), respectively (see Fig.~\ref{fig3}).
Note that in the nGBW (A) analysis, the nuclear radius $R_A$ is given by~(\ref{RA})
and free parameters
$\delta = 0.79 \pm 0.02$ and
$\pi R_p^2 = 1.55 \pm 0.02$~fm$^2$
were determined\cite{GeometryScaling-2} from experimental data on $\gamma^*A$ collisions.
In the nGBW (B) calculations,
the additional relation $R_A/R_p = A^{1/3}$ is assumed.
Despite the similar values of 
phenomenological parameters $\delta$ and $\pi R_p^2$
obtained in both our fits and nGBW (A) analysis,
we find a notable difference between the
corresponding predictions for effective nuclear gluon densities.
Moreover, the nGBW (A) and nGBW (B)
results also differ from each other.
It is explained, of course, by the 
%essentially 
distinctions in the theoretical input
used in these calculations\footnote{We successfully recover
the results\cite{nGBW}, where nGBW (B) scenario was applied.}.
In fact, while an $A$-dependence of $Q_{s,\,A}^2(x) \sim A^{1/3}$,
implemented into the nGBW (B) calculations,
is often assumed, there are approaches where 
much stronger, $Q_{s,\,A}^2(x) \sim A^{2/3}$, weaker, $Q_{s,\,A}^2(x) \sim A^{\alpha}$ with $\alpha \ll 1/3$, or 
even logarithmic $A$-dependencies are favored (see, for example,\cite{QsADependence1, QsADependence2, QsADependence3, QsADependence4, QsADependence5, nCTEQ} and references therein).
%So, we would like to stress that the transformation of the saturation scale in nucleus $Q_s\to Q_{s\,A}$
% leads to a significant nuclear modification ratio (compared to unity), and moreover, 
% it depends on the exact way such a transformation is implemented, which 
% can be seen from a comparison of the nGBW (A) and nGBW (B) scenarios: a moderate variation of
%  the parameters results in a clear difference of the integrated gluon distribution ratios. 
%  For our approach the form of the modification and the corresponding parameters come from the fit on the $F_2^A$ data.

In addition, we plot here 
nPDFs obtained by the nCTEQ Collaboration\cite{nCTEQ}
and based on the global fit to nuclear data in the framework of the DGLAP evolution scenario.
It also shows many differences with the approaches discussed above.
However, we point out that direct comparison between the effective (${\mathbf k}_T^2$-integrated)
nTMDs and conventional nPDFs can only be approximate
since these quantities are essentially different objects. 
In fact, there is no obvious resummation of any type of large logarithmic
terms in the modelled TMD gluon densities~(\ref{eq-GBWgluon}) or~(\ref{eq-input}).

\begin{figure}
	\begin{center}
		\includegraphics[width=3.9cm]{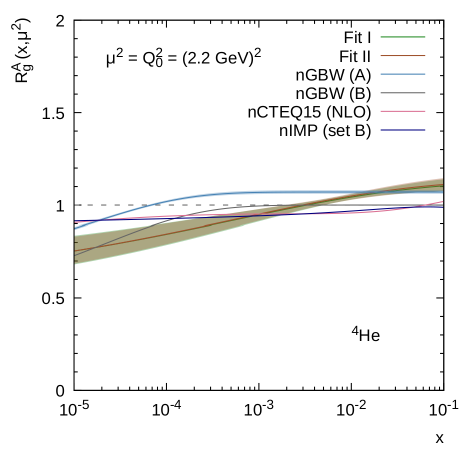}
		\includegraphics[width=3.9cm]{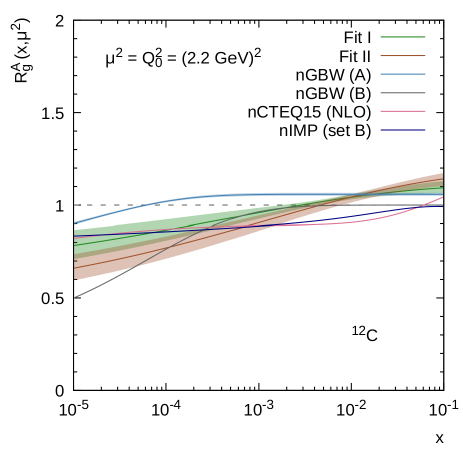}
		\includegraphics[width=3.9cm]{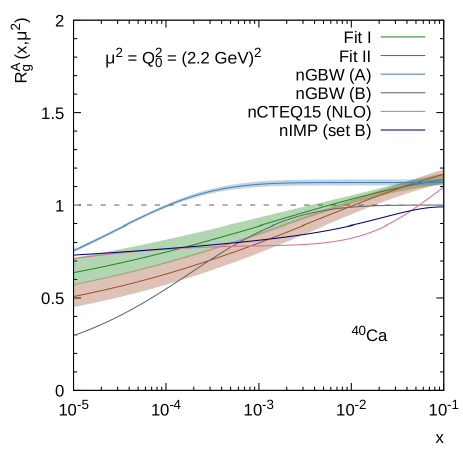}
		\includegraphics[width=3.9cm]{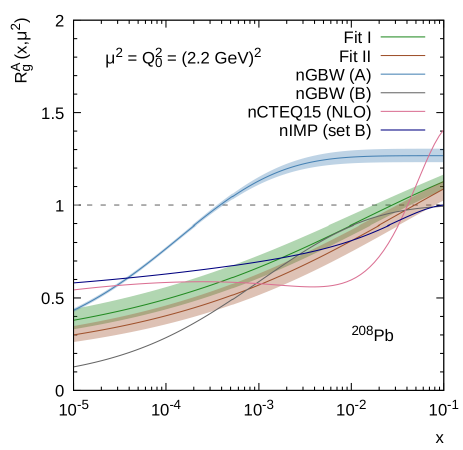} 
		\caption{Nuclear modification factor for gluon densities in some light and heavy nuclei (namely, $^4{\rm He}$, $^{12}{\rm C}$, $^{40}{\rm Ca}$ and $^{208}{\rm Pb}$)
			calculated as functions of $x$ at $\mu^2 = Q_0^2$.
			For comparison we show here corresponding results obtained by the nCTEQ\cite{nCTEQ} and nIMP\cite{nIMP} groups.}
		\label{fig4}
	\end{center}
\end{figure}

Now we turn to the discussion of the nuclear shadowing effects in the
gluon distributions at low $x$. As usual, we consider the 
nuclear modification factor for light and heavy nuclei defined as a ratio
$R^A_g(x,\mu^2) = (1/A) xg^A(x,\mu^2) / xg(x,\mu^2)$.
The results of our calculations are shown in Fig.~\ref{fig4}.
These calculations have been done at $\mu^2 = Q_0^2$,
which approximately corresponds
to the kinematical regime of %nuclear 
experiments\cite{EMC-CCuSn, EMC-CCa1, EMC-CCa2, EMC-Cu, NMC-HeCCaCCaovLi, NMC-LiC, NMC-CCaovLi, NMC-BeAlCaFeSnPbovC, NMC-SnovC, SLAC-Fe, E665-CCaPb, E665-Xe} 
performed by the EMC, NMC, SLAC and E665 Collaborations at $x < 0.1$.
Exact value of $\mu^2$ is not important here since
the scale dependence of nuclear modification factor $R^A_g(x,\mu^2)$ at low $\mu^2$
is almost negligible.
As it was expected, we observe that both our fits results in a sizeble gluon
shadowing. %which are important at small $x$.
This effect becomes more important
with decreasing of $x$ and increasing of $A$.
Nevertheless, the nuclear modification
%of gluon
in all the considered approaches shows some difference. 
So, in our calculations using Fit I it is a little weaker in comparison with
predictions of Fit II.
The shadowing magnitude obtained in the nGBW (A)/(B) calculations at low $x$
is a bit smaller/larger than both of our results, respectively.
At the same time,
predictions of Fit I and Fit II
%for gluon shadowing 
at low $x \leq 10^{-4}$ 
are typically a bit stronger for all nuclei than 
that of global analyses performed by the nCTEQ\cite{nCTEQ} or nIMP\cite{nIMP} groups
within the DGLAP framework.
Around $x \sim 10^{-3}$, our results are rather close to them.
At larger $x$, there is some gluon anti-shadowing effect, which is
observed in both Fit I and Fit II scenarios.
This effect is 
an interesting feature and
clearly seen also in the nGBW (A) and nCTEQ
calculations. In contrast, it is practically absent in the nGBW (B) and nIMP analyses.
Such difference in predictions of these approaches is notable and shows the importance 
of studying the gluon nuclear modification factor
in future experiments.
In the applied approach, this factor
depends on the exact way of the transformation of the saturation 
scale in nucleus (see\cite{GeometryScaling-2}). % $Q_s^2(x)\to Q_{s\,A}^2(x)$. %, which is widely discussed in the literature\cite{QsADependence1, QsADependence2, QsADependence3, QsADependence4, QsADependence5}.
We stress that in our fits the form of such modification and 
corresponding parameters come from the best description of the data
on the nuclear structure functions $F_2^A(x,Q^2)$.
Indeed, more experimental data are needed to distinguish
between the different nPDF models.

To conclude, 
the main outcome of the present study is new TMD gluon densities in a nucleus.
We extend the previously proposed TMD gluon density in a proton %proposed earlier\cite{?}
to nuclei using the well established empirical property of geometric scaling.
It modifies the gluon saturation scale $Q_s(x)$ in a way to incorporate nuclear medium effects 
leading to shadowing of gluon distributions at small $x$.
It is important that, in contrast with other studies, the input gluon density in a proton applied in our analysis
is able to reproduce a number of HERA and LHC data,
that is an advantage of our consideration.
The parameters of the scale modification have been obtained from a fit to various data on 
ratios $F_2^{A}(x, Q^2)/F_2^{A^\prime}(x, Q^2)$
for several nuclear targets %$A$ and $A^\prime$
at low $x < 0.1$.
The latter have been calculated using the color dipole formalism.
Two fits relying on different formulae for the nuclear radius have been performed and 
both of them provide reasonable description of the data.
So, we have shown that one can include the shadowing effect in nuclei collisions by the modification of  
the saturation scale
and demonstrate that the developed approach %\cite{LLM-2022, LLM-2024}
can be applied in the phenomenological studies in the shadowing region
within the dipole formalism.
Moreover, our fits could be used also
as an initial condition for the subsequent non-collinear QCD evolution.
It is important for future studies of lepton-nucleus, proton-nucleus and 
nucleus-nucleus interactions at modern and future colliders, 
%within the TMD-based approaches, 
where 
the gluon saturation dynamics could be examined directly.

{\sl Acknowledgements.} We thank S.P.~Baranov and H.~Jung for their
interest, very important comments and remarks.
This research has been carried out at the expense of the Russian Science Foundation grant No.~25-22-00066, https://rscf.ru/en/project/25-22-00066/.

\bibliography{nTMD-1}

\end{document}